\documentstyle{l-aa}

\begin{document}

   \thesaurus{12         
              (12.04.2;  
               12.07.1;
               13.18.1)} 
   \title{Statistical excess of foreground galaxies around 
high$-z$ radiogalaxies\thanks{Based
on observations made with the 2.2m Telescope at Calar Alto}}

   \author{N. Ben\' \i tez\inst{1}\inst{,2}, 
E. Mart\' \i nez-Gonz\'alez\inst{1} \and 
J.M. Martin-Mirones\inst{1}\inst{,2}}

   \offprints{N. Ben\'\i tez}

   \institute{IFCA, CSIC-UC,
              Fac. de Ciencias, Avda. Los Castros s/n,
              39005 Santander, Spain
   \and
Dpto. de F\' \i sica Moderna, UC,
              Fac. de Ciencias, Avda. Los Castros s/n,
              39005 Santander, Spain}

   \date{}

   \maketitle

   \begin{abstract}$K-$band imaging of a sample of 
21 radiogalaxies with $z \sim 1.5$ reveals the existence of a statistical 
association of foreground galaxies with the positions of the radiosources.
The excess is detected within a 1' radius at a high significance 
level ($>99.8\%$). $K$-band light is a good tracer of stellar mass, 
so this result indicates the existence of an association between foreground 
mass perturbations and background radiosources, as expected from the 
magnification bias effect, and confirms previous results obtained for 
other high$-z$ radio-loud AGN. 
      \keywords{radiogalaxies--
                gravitational lensing--
                magnification bias}
   \end{abstract}

%

\section{Introduction}

The study of statistical associations between foreground galaxies and  
background AGNs has a long and controversial history. 
Gravitational lensing seems to provide the best explanation for this
phenomenon (see Schneider et al 1992, Narayan and Bartelmann 1996, Schneider 1996 
for recent reviews of both theoretical and observational aspects).

The density of a 
population of flux-limited background sources 
(e.g. AGNs), magnified by a factor $\mu$ by a mass 
perturbation 
is changed in two ways. As the angular dimensions of the lensed 
zone are expanded by a factor $\mu$, the physical size of a 
region observed through a fixed angular aperture will be smaller 
than in the absence of the lens. This causes a decrease in the 
AGN surface density equal to $\mu^{-1}$.
On the other hand, the lens will magnify faint sources (which would not 
have been detected otherwise) into the sample and increase the number
of detected AGN. If the slope of the AGN number-counts cumulative 
distribution is steep enough, the latter effect would dominate over the 
former and there would be a net excess of AGN behind the lens. 
The value of the density enhancement $q$ of these sources 
around foreground galaxies would be $q=\mu^{\alpha_e-1}$, where 
$\alpha_e$ is the effective slope of the background source cumulative 
number counts ( Borgeest et al. 1991).

Observations show that there seems to be 
positive correlations between foreground galaxies
and high$-z$ radio-loud quasar samples (Hintzen et al. 1991, 
Bartelmann \& Schneider 1994, 
Ben\'\i tez \& Mart\'\i nez-Gonz\'alez 1995, etc.), 
or heterogeneous samples which contain a considerable fraction of 
radio-loud quasars ( Tyson 1986, Thomas et al. 1995). 
The results for radio-quiet QSOs are confuse and often
even contradictory. Although at small scales there are both 
positive and negative results (see Narayan 1992), at large scales 
there are no detections of radio-quiet QSO-foreground galaxy positive 
correlations but evidences of anticorrelations 
(Boyle et al. 1988, Ben\'\i tez \& 
Mart\'\i nez-Gonz\'alez 1997). 

 This last fact cannot be explained in terms of gravitational lensing alone: 
although stronger correlations are predicted for radio-loud samples 
due to the existence of a double magnification bias effect (Borgeest
et al. 1991), the correlations should also be observed for radio-quiet 
QSO samples. These results, however, may be understood if we consider
small amounts of dust associated with the lensing masses (Ben\'\i tez \& 
Mart\'\i nez-Gonz\'alez 1997), a phenomenon which may be called 'dirty' 
or 'dusty' lensing. In that case, and taking into account the double 
magnification bias, for a radio-loud sample the density enhancement 
$q_r$ would be $q_r=\mu^{\alpha_r+\alpha_o-1}e^{-\alpha_o\tau}$, 
whereas for the radio-quiet sample $q_o=\mu^{\alpha_o-1}e^{-\alpha_o\tau}$, 
where $\alpha_o$ and $\alpha_r$ are respectively the cumulative 
number-counts slope in the optical ($N(<m)\propto 10^{0.4\alpha_o m}$) 
and the radio bands ($N(>S) \propto S^{-\alpha_r}$) respectively 
and $\tau$ is the optical depth by dust present in the lensing 
matter. We have assumed that the absorption is much larger in the 
optical than in radio, and that the radio flux is approximately 
independent from the optical magnitude. Then, if e.g. 
$\mu e^{-\tau} \approx 1$, we have that 
$q_r \approx \mu^{\alpha_r-1}$, independently of the 
number-counts slope in the optical $\alpha_o$, and $q_o 
\approx \mu^{-1}$ for the radio-quiet QSOs. 

It is thus specially interesting to determine if the 
associations with foreground galaxies also occur for other
radio-loud AGN samples which have very different optical 
properties from the quasars, as the radiogalaxies.
In Ben\'\i tez et al. 1995 (B95), an excess of apparently foreground
galaxies was found around a sample of 5 radiogalaxies with $1<z<2$. 
In this paper we extend our study to a larger sample (21 radiogalaxies), 
and perform it in the $K$ band. This band is a good tracer of old stellar 
populations and makes easier the identification of foreground 
mass concentrations. 

This paper is structured as follows: In Sec. 2 we report the observation
and data reduction procedures. In Sec. 3 we describe the results 
and the statistics. Sec. 4 discusses our main results and conclusions.

%

\section{Observations and data reduction}

Our preliminary sample was formed by all the 
radiogalaxies from the 3C catalogue with $0.95 < z < 2$. 
We observed 17 sources\footnote{
 3C013, 3C065, 3C068, 3C124, 3C173, 3C184, 3C194, 3C208, 
 3C210, 3C256, 3C267, 3C280, 3C294, 3C322, 3C324, 3C326, 
 3C356} from this catalogue picking them in function
of the availability of their coordinates and completed the final 
sample with four radiogalaxies with similar characteristics obtained from 
the literature: 0956+475, 1129+35, 1141+354 and 1230+349. As far as 
we know there is no bias whatsoever towards the presence of low redshift 
objects around our sample. 

The observations were carried out at the CAHA 2.2m Telescope at 
Calar Alto with the camera MAGIC on the nights of 1995 January 16-19 
and May 16-17. MAGIC uses a $256\times256$ NICMOS3 HgCdTe detector array. 
For the purposes
of our investigation it is essential to have a wide field in order to 
determine accurately the background density, so we used the configuration 
which provided the 
largest pixel scale, $1.61$ arcsec/pixel, which gives a 
$\approx 6.9\times 6.9$ arcmin$^2$ field. 

We typically took 27 exposures of 60 sec. for each field. The telescope was 
nodded among nine positions forming a 20 arcsec square grid. 
Both in the january and may runs the weather ranged from bad to awful, 
and some of the exposures had to be rejected due to clouds. 
The images have been 
reduced in the standard way for near IR observations.
A first pass reduction involved forming a flat field for each frame by median 
combining the closest frames with a sigma-clipping algorithm.
After identifying the most extended and saturated objects, we eliminated
them from the raw frames and repeated the procedure. The final, averaged 
$K$ images were trimmed and only contain the intersection zone
which was observed in all the exposures. The median p.s.f. 
was $\approx 2.5$ arcsec. Although the nights were not photometric,  
the way in which the analysis has been performed ensures that our 
main results are insensitive to photometric uncertainties. 

The package SEXtractor (Bertin and Arnouts 1996) was used for image 
detection and classification. We accurately determined
the p.s.f. of each field running the routine SEXseeing 
several times until it converged to a stable value and then 
performed the detection after smoothing the frame with a gaussian close 
to the p.s.f. value. 

The star-galaxy classification algorithm correctly identifies 
the brightest, saturated stars. At fainter magnitudes, with our
pixel scale it is virtually impossible to distinguish between stars and
galaxies. In any case, any star entering our sample would just 
tend to dilute any possible excess and lower the significance of our
results.  So we form our galaxy sample by taking all the objects 
to which SEXtractor assigns a stellar index smaller than 0.9, 
which excludes the more conspicuous bright stars. We have checked that 
the final results are insensitive to variations in the stellar index threshold.
We set the completeness limit at the magnitude at which the 
detected galaxies typically have a photometric error of $0.3$. This usually 
corresponds to $K\approx 18-18.5$. 

The radiogalaxy has been detected in 11 of the fields, but there 
are several instances in which either it is fainter than the 
completion limit (5) or the presence of a bright object 
close to its position makes impossible either identification or 
photometry (5). 
When needed we have established its location in the frame 
with an accuracy of $\approx 2$ arcsec with the help of reference 
charts from the Digital Sky Survey, and obtained its magnitude
from the literature or estimated it with the $K-z$ relationship
(McCarthy 1993 and references therein)  

The distribution of 'truncated' or incomplete objects in the fields 
shows that there is a 10 pixel wide zone where SEXtractor 
loses bright objects because they are too close to the border of the 
detector. We have excluded this region from our catalogues and also a 
10 pixel radius circle around the 
radiogalaxy to avoid merged objects. The 'useful' surface thus 
defined contains 1944 galaxies and covers a surface of 645.4 arcmin $^2$. 

\section{Statistical analysis and results}

To check for the presence of associations due to the magnification bias
effect, we have to select a foreground galaxy catalogue and try to avoid
the inclusion of any galaxies clustering with the radiogalaxy at 
its same redshift. We can make use of the fact that the high$-z$ 
3C radiogalaxies are representative of some of the brightest 
known early type galaxies, so they are very likely to 
be more luminous than any other object at the same redshift, and 
redder than almost any foreground galaxy. 
Therefore, any object which is significantly brighter in $K$ than 
the radiogalaxy is almost certainly in the foreground. 

\begin{table}
      \caption{Number of objects in the central 1' region}
         \label{Tab1}
\begin{flushleft}
\begin{tabular}{llll}
\hline\noalign{\smallskip}
$K-K_{rg}$ & $n_f$ & $n_e$\\
\noalign{\smallskip}
\hline\noalign{\smallskip}
$< -5.0$ &0   & 0.09  \\
$< -4.5$ &0   & 0.19  \\
$< -4.0$ &1   & 0.28  \\
$< -3.5$ &2   & 0.75  \\
$< -3.0$ &7   & 2.16  \\
$< -2.5$ &13  & 5.63  \\
$< -2.0$ &22  & 11.04 \\
$< -1.5$ &31  & 19.02 \\
$< -1.0$ &55  & 34.34 \\
$< -0.5$ &72  & 55.91 \\
$<  0.0$ &111 & 90.48 \\
$<  0.5$ &138 & 117.98\\
$<  1.0$ &165 & 144.38\\
$<  1.5$ &190 & 170.88\\
$<  2.0$ &205 & 183.48\\
$<  2.5$ &207 & 187.69\\
\noalign{\smallskip}
\hline
\end{tabular}
\end{flushleft}
\end{table}

We employ a statistical method similar to the one discussed and
applied in B95. The number of galaxies found in a central
region around the radiogalaxies $n_f$ is compared with the expectation
$n_e$ determined from the average density on the whole 'superfield'
(including the central region). To compare with previous results for
radiogalaxies (B95), a 1' radius circle was chosen. 
In Fig. 1 we 
have plotted histograms for $n_f$ ( continuous line) and $n_e$
(dashed line) binned in $0.5$ mag intervals of $K-K_{rg}$, the
difference between each galaxy magnitude and the magnitude of the
radiogalaxy in its field. An excess of galaxies is clearly seen in the 
magnitude bins which are brighter than the radiogalaxies. 
In Table 1
we have listed $n_f$ and $n_e$ for galaxies brighter than a given
threshold in $K-K_{rg}$.  In order to have enough galaxies for our
statistical analysis we take the objects which have $K<K_{rg}-1$, 
which is enough to ensure that the objects are brighter than
the radiogalaxies. 
We find 55 galaxies within the central 1' region which are at least 
1 mag brighter than the corresponding radiogalaxy, against an expectation 
of 34.34 ($q=1.60$). The density enhancement would be even stronger 
if we set a brighter threshold ($q=1.99$ for $K<K_{rg}-2$; 
$q=3.24$ for $K<K_{rg}-3$) but the significance levels would 
be lower due to the smaller total number of galaxies in the samples.

\begin{figure}
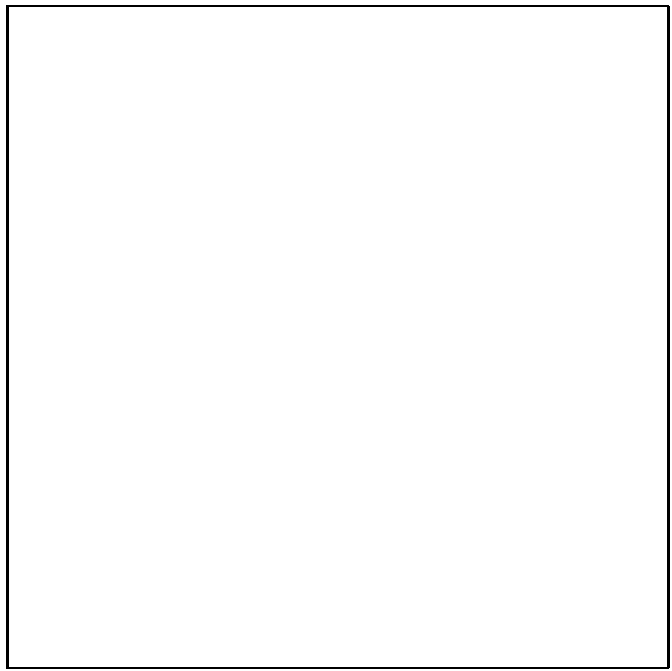

\picplace{8.8cm}
\caption[]{Distribution of galaxies as a function of $K-K_{rg}$}
\end{figure}
How significant is this result? If we placed $N_i$ points at random 
within a field of surface $\Omega_{Ti}$, the number of points 
$n_i$ found within a central region of fixed surface $\Omega_C$ would be 
binomially distributed with average $<n_i>=N_i p_i$ and variance 
$\sigma n_{ri}^2=N_i p_i(1-p_i)$, where $p_i=\Omega_C/\Omega_{Ti}$. 
In our case $p_i \approx 0.1$, which gives a $5\%$ difference 
with respect to Poisson that would vanish in the limit when $\Omega_C$ 
is much smaller than $\Omega_{Ti}$. 

If the results of 21 different independent fields were added up together, 
the resulting distribution would be approximately gaussian 
with an average $<n>=\sum N_i p_i$ and a variance $\sigma_r^2
=\sum N_i p_i(1-p_i)$. 
We expect that the variance for a real galaxy field 
$\sigma_{gi}^2$ will be larger than that of a random 
distribution $\sigma_{ri}^2$ because of galaxy clustering (Peebles 1980):
\begin{equation}
\sigma_{gi}^2\approx N_i p_i (1-p_i)+
{(N_i p_i)^2\over \Omega_C^2}\int_{\Omega_C}
 w_i(\theta_{12})d\Omega_1 d\Omega_2 
\end{equation}
Baugh et al. 1996 have measured an amplitude $A_{15}\approx 0.03$ for the 
correlation function $w(\theta) =A\theta^{-0.8}$ of $K<15$ galaxies. 
As it is well known, the correlation function approximately scales 
with depth as $w(\theta) =D^{-1}w(\theta D)$, where 
$D=10^{0.2(m-M^*)-5}$Mpc (Peebles, 1980). 
Thus the amplitude will change with the limiting magnitude as
$A_K \approx 
A_{15} 10^{0.36(15-K)}$.
After calculating the integral of the correlation function 
with a Monte Carlo taking into account the shape of our region we 
found that the total dispersion combining the 21 separate fields 
would be:
\begin{equation}
\sigma_g^2 \approx \sum N_i p_i(1-p_i) + 35.15 \sum A_i N_i^2 p^2_i 
\end{equation}
where $A_i$ is the amplitude of the correlation function corresponding
to each field. Thus the 'expected' variance in
the number of galaxies within the central region should be 
$\sigma_g^2 \approx 31.02 + 18.55$.
If we measure empirically the rms from our data we obtain 
a lower value, $\sigma \approx 5.6$, very close to the expectation 
for a random galaxy distribution. However, this empirical estimation is 
rather uncertain, because our total surface is only $\approx 10$ 
times larger than the central region, and thus the amount of
independent data is very limited. For this reason, we take as a 
conservative estimate of the true dispersion the theoretical upper limit, 
$\sigma_g \approx 7$, which yields a significance of at least 
$2.95\sigma$, equivalent to a $99.8\%$ significance level.

Incidentally, if we consider the four fields which are common to our 
sample and to B95, namely 0956+475, 3C256, 3C294 and 3C324 (1217+36 was 
excluded from the present sample because it was found to be stellar-like 
in B95), we found 12 foreground galaxies in the central region against an 
expectation of 7.26, a $q=1.65$ excess close to the obtained 
for the full sample. 

To further check the above mentioned results, we have measured the central 
density of objects which are {\sl fainter} than our foreground sample. 
The sample thus defined contains 1587 galaxies. No excess is detected: 
there are 152 galaxies in the central box against an expectation of 153.3. 

\section{Discussion and conclusions}

There are few deliberate searches for magnification bias effects 
around high$-z$ radiogalaxies in the literature. Besides, the 
comparison of our results with other works is complicated by the 
fact that as far as we know, this is the first such analysis 
done in the near IR. 

Our present result clearly confirms the excess found in B95. 
However, a bootstrap analysis shows that subsamples of 4 fields 
chosen randomly from our sample yield overdensities ranging from 
$1.14 < q < 2.09$ ($68\%$ confidence limits). If the average density 
enhancement $q$ were slightly lower, which could be the case for 
other galaxy or AGN samples, values of $q\approx 1$ could have been 
obtained in a non-negligible number of cases. This should be taken into account 
when working with very small samples, as in Fried 1996. 

Hammer and Le Fevre (1990) claim that there are 
nine times more high$-z$ 3CR galaxies within five arcsec of $R\leq 21$ 
galaxies than expected from a random distribution. 
Although we cannot check the excess at  
those scales because of our pixel scale, our results confirm 
that samples as the 3CR radio sources may be considerably affected
by gravitational lensing.

In a recent paper, Roettgering et al. (1996) do not find any excess of bright 
galaxies $R<21.5$ around a sample of 11 ultra-steep-spectrum radio 
galaxies with $ z>1.8$. However, the scale of their field is smaller 
than ours, $r < 100''$. 
The number of foreground galaxies found within $100''$ of our 
radiogalaxies is 129. Thus, the expectation within the central
$60''$ would be $44.27 \pm 7.9$ ( the rms has been estimated as above). 
We found 55, that is a $q=1.24 \pm 0.18$ overdensity with 
a $1.35\sigma$ significance level. For a sample of 11 galaxies 
the significance would have been just $\approx 1\sigma$, against
the almost $3\sigma$ that we obtain in our larger fields.
In any case, this comparison is not very conclusive, as far as 
neither the filter, nor the radiogalaxy type or redshift distribution 
are the same. 

  We have determined the existence of an statistical excess of 
foreground galaxies around a sample of background, high redshift 
radiosources. The way in which the foreground sample galaxies
have been selected almost excludes the possibility of their being 
physically associated with the radiogalaxies. The only consistent 
explanation~---at least qualitatively---for this fact is the magnification 
bias caused by gravitational lensing by the dark matter perturbations 
traced by the foreground objects. 

The small size of the sample 
precludes a detailed, quantitative study of the form of the 
radiogalaxy-galaxy correlation function. It would be therefore 
desirable to use larger, well understood radiogalaxy samples
with accurately defined properties to obtain a robust 
estimate of $q$ which could be compared with theoretical estimations. 

\begin{acknowledgements}
We thank the referee, Peter Schneider, for his valuable comments.
The 2.2m telescope is operated by the Max-Planck-Institut f\"ur Astronomie
at the Centro Astron\'omico Hispano-Alem\'an in Calar Alto (Almer\'\i a, 
Spain). 
The Digitized Sky Survey was produced at the 
Space Telescope Science Institute (ST ScI) under U. S. Government 
grant NAG W-2166. NB and EMG acknowledge financial support from 
the Spanish DGICYT, project PB95-0041. NB acknowledges a Spanish 
M.E.C. Ph.D. scholarship.  
\end{acknowledgements}

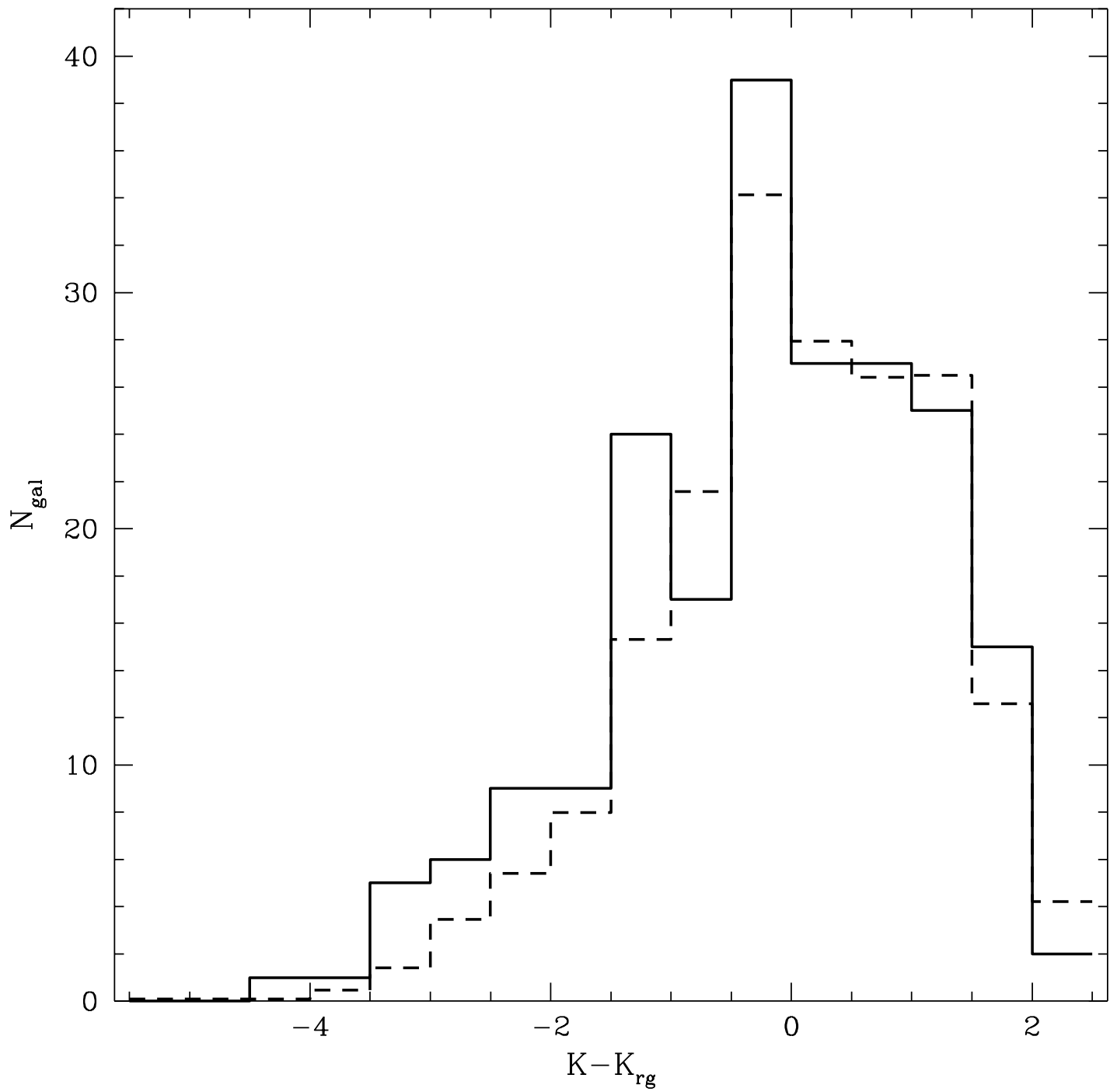
\end{document}